\documentclass[aps,amssymb,amsfonts,article,prl,floatfix,showpacs,superscriptaddress]{revtex4}
\usepackage{amssymb}
\usepackage{amsmath}
\usepackage{amsfonts}
\usepackage{amsopn}
\usepackage{graphics}
\usepackage{graphicx}
\usepackage{bm}
\usepackage{subfigure}
\usepackage{color}

\usepackage[utf8]{inputenc}
\usepackage{siunitx}
\usepackage{cleveref}

\begin{document}

\title{Energy conservation in self-phase modulation}

\author{P. B\'ejot}
\affiliation{Group of Applied Physics, University of Geneva, GAP, 22 chemin de Pinchat, 1211 Geneva 4, Switzerland}

\author{J. Kasparian}
\email{jerome.kasparian@unige.ch}
\affiliation{Group of Applied Physics, University of Geneva, GAP, 22 chemin de Pinchat, 1211 Geneva 4, Switzerland}
\affiliation{Institute for Environmental Sciences, University of Geneva, bd Carl Vogt  66, 1211 Geneva 4} 

\date{\today}
\begin{abstract}
Spectral broadening of ultrashort laser pulses is simultaneously described by either self-phase modulation (SPM) or four-wave mixing (FWM). The latter implies the instantaneous conservation of both the photon number and energy, while the former describes a time-dependent frequency shift, implying a violation of the energy conservation if the number of photons is to be conserved in each time slice. 
We resolve this paradox by considering the transient energy storage in the propagation medium, that can be calculated in the SPM formalism via the dephasing between the incident pulse and the medium polarization leading to an effective imaginary part in the third-order susceptibility. In parallel, considering the temporal variation of the incident intensity in FWM offsets the instantaneous frequency.
\end{abstract}
\maketitle

\section{Introduction} 
Non-linear optics has been first investigated in condensed matter. The spectacular phenomenon of supercontinuum generation was observed as early as 1970 by Alfano and Shapiro~\cite{AlfanoS70b,AlfanoS70a}, who identified four-wave mixing (FWM)~\cite{AlfanoS70b} and self-phase modulation (SPM)~\cite{AlfanoS70a} as the origin of this wide spectral broadening~\cite{Zheltikov2002,DudleyRMP,KnightS07,Alfano06pXX,BoydXX,He_Liu,LoudonXX,Agrawal,Shen}. The advent of chirped pulse amplification~\cite{StricklandM85} in $1985$ allowed even more efficient spectral broadening, as well as its observation in gases including atmospheric pressure air.

FWM describes spectral broadening as the interaction of plane monochromatic waves. Two photons at frequencies $\omega_1$ and $\omega_2$ interact through the $\chi^{(3)}$ susceptibility  to generate two photons at $\omega_3$ and $\omega_4$. The energy conservation imposes $\omega_1+\omega_2=\omega_3+\omega_4$, so that the spectrum  should remain symmetrical after broadening via FWM, and the number of photons is conserved at any time.

The same spectral broadening is alternatively described by SPM as a deformation of the pulse. After propagating over a distance $z$, the carrier wave experiences a time-dependent frequency shift

\begin{equation}
\Delta\omega(t)=-k_0 z n_2 I' (t) z,
\label{SPM}
\end{equation}
where the prime denotes the temporal derivation, $k_0=n_0\omega/c$, $n_0$ and $n_2={3\chi^{(3)}}/{4n_0 \epsilon_0 c}$ are the linear and non-linear refractive indices, respectively, $\omega$ the angular frequency of the incident pulse, $I$ its intensity, and $c$ the velocity of light. 
SPM intrinsically induces an asymmetric time-dependent frequency shift. In most common media like air or glass $n_2>0$, so that the leading edge of the pulse is red-shifted while its trailing edge is blue-shifted. Such time-dependent frequency shift is incompatible with the simultaneous conservation of the pulse energy and of the number of photons.

In this Letter, we address this paradox. By taking into account the dephasing induced by SPM between the incident electric field and the polarization of the propagation medium, we show that the latter transiently stores energy, which restores energy conservation. Furthermore, considering the fast variation of the pulse intensity in the FWM formalism yields the asymmetric transient frequency shifts that are usually described in the SPM formalism.

\section{Discussion}
\subsection{Energy conservation in SPM.}
We consider a pulse described within the slowly-varying envelope approximation (SVEA), so that the electric field expresses as $E(t)=E_0(t)(e^{-\mathrm{i}\omega t}+\mathrm{c.c.})$.  It induces a third-order polarization 
\begin{equation}
P^{(3)}(t) = \epsilon_0 \chi^{(3)}E(t)^3 \label{eq:P3}
\end{equation}
$\epsilon_0$ being the permittivity of vacuum.
If we neglect absorption and dispersion, the evolution of this third-order polarization in a medium with an eigenfrequency $\omega_\mathrm{e}$ can be written in a perturbative approach \cite{EttoumiPKW10}:

\begin{equation}
\frac{\mathrm{d}^2 P^{(3)}}{\mathrm{d}t^2} + {\omega_\mathrm{e}}^2 P^{(3)} = N e Q^{(3)} {P^{(1)}}^3(t)
\label{P3}
\end{equation}
where $P^{(1)} = \chi^{(1)}E$ is the elastic polarizability,  $\chi^{(1)}$ the first-order susceptibility, $E$ the electric field, $N$ the local density of electrons, $-e$ their charge, $m$ their mass, and
\begin{equation}
Q^{(3)}=\frac{N^3 e^4}{m\epsilon_0^3}\frac{\chi^{(3)}}{{\chi^{(1)}}^4}
\end{equation}
Expanding the first-order polarization $P^{(1)}(t)=P_0^{(1)}(t)\left(e^{i\left(\omega t - k z - \phi^{(1)}\right)} + \mathrm{c.c.}\right)$ in  the right term of Equation (\ref{P3}), we get

\begin{equation}
\frac{\mathrm{d}^2 P^{(3)}}{\mathrm{d}t^2} + {\omega_\mathrm{e}}^2 P^{(3)} = N e Q^{(3)}{P_0^{(1)}}^3 
\times \left( e^{3i\left(\omega t - kz - \phi^{(1)}\right)} + 3e^{i\left(\omega t - kz - \phi^{(1)}\right)} + \mathrm{c.c.}\right)
\label{P3_bis}
\end{equation}

We decompose the polarization $P^{(3)}$ into its components  $P^\mathrm{SPM}$ and $P^\mathrm{TH}$, respectively oscillating at the fundamental and third harmonic (TH) frequencies $\omega$ and $3\omega$:

\begin{equation}
P^{(3)}(t)=P_0^\mathrm{TH}(t)\left(e^{i\left(3(\omega t-kz)-\phi^\mathrm{TH}\right)}+\mathrm{c.c.}\right)
+P_0^\mathrm{SPM}(t)\left(e^{i\left(\omega t-kz-\phi^\mathrm{SPM}\right)} + \mathrm{c.c.}\right)
\label{eq:defPTH_PSPM}
\end{equation}

Here, $\phi^\mathrm{TH}$ and $\phi^\mathrm{SPM}$ are the dephasings of the polarization relative to the electric field for each spectral component. They are chosen such that $P_0^\mathrm{TH}$ and $P_0^\mathrm{SPM}$ are real. We focus on the terms oscillating at the fundamental frequency $\omega$, consider the SVEA approximation, identify the real and imaginary parts of $P^\mathrm{SPM}$, and calculate $\phi^{(1)}$ from the linear polarization equation~(\ref{eqn:P_lin}) \cite{EttoumiPKW10}. 

The SVEA implies ${\mathrm{d}\phi^\mathrm{SPM}} / {\mathrm{d}t} \ll \omega$ and ${\mathrm{d}^2P_0^\mathrm{SPM}} / {\mathrm{d}t^2} \ll \omega^2 P_0^\mathrm{SPM}(t)$, so that
the terms at frequency $\omega$ in Equation (\ref{P3_bis}) rewrite:

\begin{widetext}
\begin{multline}
\left(\omega_\mathrm{e}^2-\left(\omega-\frac{\mathrm{d}\phi^\mathrm{SPM}}{\mathrm{d}t}\right)^2\right) P_0^\mathrm{SPM}(t)
+ 2i\left(\omega-\frac{\mathrm{d}\phi^\mathrm{SPM}}{\mathrm{d}t}\right)\frac{\mathrm{d}P_0^\mathrm{SPM}}{\mathrm{d}t} 
+\frac{\mathrm{d}^2P_0^\mathrm{SPM}}{\mathrm{d}t^2}
= 3N e Q^{(3)} {P_0^{(1)}}^3(t) \left(e^{i\left(\phi^\mathrm{SPM}-\phi^{(1)}\right)} +\mathrm{c.c.} \right)\\
=3N e Q^{(3)}\left(\epsilon_0\chi^{(1)}E_0\right)^3 \left(e^{i\left(\phi^\mathrm{SPM}-\phi^{(1)}\right)}+\mathrm{c.c.} \right)
\end{multline}
\end{widetext}

Identifying the components parallel and orthogonal to $P_0^\mathrm{SPM}$ in the complex plane yields

\begin{equation}
\left(\omega_\mathrm{e}^2-\omega^2\right) P_0^\mathrm{SPM}(t) = 6N e Q^{(3)}\left(\epsilon_0\chi^{(1)}E_0\right)^3 
\times \cos\left(\phi^\mathrm{SPM}-\phi^{(1)}\right) \label{eqn:P_SPM_ER}
\end{equation}

\begin{equation}
2\omega\frac{\mathrm{d}P_0^\mathrm{SPM}}{\mathrm{d}t} = 6N e Q^{(3)}\left(\epsilon_0\chi^{(1)}E_0\right)^3 
\times \sin\left(\phi^\mathrm{SPM}-\phi^{(1)}\right) \label{eqn:P_SPM_EI}
\end{equation}

which combine into

\begin{eqnarray}
\phi^\mathrm{SPM}-\phi^{(1)} &\approx& \tan \left(\phi^\mathrm{SPM}-\phi^{(1)}\right) \\
&=&\frac{2\omega}{\omega_\mathrm{e}^2-\omega^2}\frac{\mathrm{d}P_0^\mathrm{SPM}}{P_0^\mathrm{SPM}(t)dt} \\
&=&\frac{6\omega}{\omega_\mathrm{e}^2-\omega^2}\frac{\mathrm{d}E_0}{E_0(t)\mathrm{d}t} \\
&=&\frac{3\omega}{\omega_\mathrm{e}^2-\omega^2}\frac{\mathrm{d}I}{I(t)\mathrm{d}t}
\end{eqnarray}

$\phi^{(1)}$ can be estimated with a similar derivation starting from the linear propagation equation \cite{EttoumiPKW10}:

\begin{equation}
\displaystyle{\frac{\mathrm{d}^2 P^{(1)}}{\mathrm{d}t^2} + {\omega}^2 P^{(1)} = \frac{N e^2}{m} E(t)}
\label{eqn:P_lin}
\end{equation}
resulting in
\begin{equation}
\phi^{(1)}=\frac{\omega}{\omega_\mathrm{e}^2-\omega^2} \frac{\mathrm{d}I}{I(t)\mathrm{d}t}
\end{equation}
which finally yields

\begin{equation}
\phi^\mathrm{SPM}=4\phi^{(1)}
=\frac{4\omega}{\omega_\mathrm{e}^2-\omega^2} \frac{\mathrm{d}I}{I(t)\mathrm{d}t}
\label{phi_spm}
\end{equation}

Far from resonance ($\omega \ll \omega_\textrm{e}$), $\phi^\mathrm{SPM} \approx {4\omega} /{\omega_\textrm{e}^2T} \ll 1$, $T$ being the pulse duration. For example, in air $\omega_\mathrm{e} \approx$ 80 nm \cite{ZhangLW2008} so that for typical experiments with a 100~fs pulse centered at 800~nm, $\phi^\mathrm{SPM} \approx 10^{-4}$. Therefore, the supplementary self-phase modulation induced by this dephasing is fully negligible. 

However, the dephasing $\phi^\mathrm{SPM}$ between the driving field $E$ and the resulting polarization $P^\mathrm{SPM}$ implies an energy transfer between the incident electric field and the propagation medium. The instantaneous power per unit volume transferred to the propagation medium amounts to
\begin{equation}
\mathcal{P}^\mathrm{SPM}(t) = E(t)\frac{\mathrm{d}P^\mathrm{SPM}}{\mathrm{d}t}
\label{Eq:P}
\end{equation}
where the value of ${\mathrm{d}P_0^\mathrm{SPM}} / {\mathrm{d}t}$ is defined in Eq.~(\ref{eqn:P_SPM_EI}), so that
\begin{widetext}
\begin{eqnarray}
\mathcal{P}^\mathrm{SPM}(t)
&=&2E_0(t)\cos\left(\omega t-kz\right)\left(-2\omega P_0^\mathrm{SPM}(t) \sin\left(\omega t-kz-\phi^\mathrm{SPM}\right)+2\frac{\mathrm{d}P_0^\mathrm{SPM}}{\mathrm{d}t}\cos\left(\omega t-kz-\phi^\mathrm{SPM}\right) \right) 
\end{eqnarray}

Averaging over one or a few optical cycles yields
\begin{eqnarray}
\langle\mathcal{P}^\mathrm{SPM}\rangle (t)&=&-4\omega E_0P_0^\mathrm{SPM}\frac{\omega}{2\pi}\int_0^{\frac{2\pi}{\omega}}{\cos(\omega t)\sin\left(\omega t-\phi^\mathrm{SPM}\right)\mathrm{d}t} + 4E_0\frac{\mathrm{d}P_0^\mathrm{SPM}}{\mathrm{d}t}\frac{\omega}{2\pi}\int_0^{\frac{2\pi}{\omega}}{\cos(\omega t)\cos\left(\omega t-\phi^\mathrm{SPM}\right)\mathrm{d}t} \\ 
&=&4\omega E_0P_0^\mathrm{SPM}\frac{\sin\phi^\mathrm{SPM}}{2}+ 4E_0\frac{\mathrm{d}P_0^\mathrm{SPM}}{\mathrm{d}t}\frac{\cos\phi^\mathrm{SPM}}{2}
\end{eqnarray}

\end{widetext}

Plugging the values of $P_0^\mathrm{SPM}$ and $\frac{\mathrm{d}P_0^\mathrm{SPM}}{\mathrm{d}t}$ from Equations~(\ref{eqn:P_SPM_ER}) and (\ref{eqn:P_SPM_EI}), 
we obtain

\begin{eqnarray}
\langle\mathcal{P}^\mathrm{SPM}\rangle(t)
&\approx&6\epsilon_0E_0^4 
\left[\omega \chi^{(3)}_{\mathrm{cw}} \sin\phi^\mathrm{SPM}+\chi^{(3)}_{\mathrm{cw}}\frac{\omega_\mathrm{e}^2-\omega^2}{2\omega} \sin\left(\phi^\mathrm{SPM}-\phi^{(1)}\right) \right]  \\
&\approx&6\epsilon_0E_0^4  \chi^{(3)}_{\mathrm{cw}}\phi^{(1)}
\left(4\omega+ 3 \frac{\omega_\mathrm{e}^2-\omega^2}{2\omega} \right)  \\
&=&3\epsilon_0 \chi^{(3)}_{\mathrm{cw}} E_0^4 \phi^{(1)} \left(\frac{3\omega_\mathrm{e}^2+5\omega^2}{\omega} \right)
\end{eqnarray}

Introducing the relations $I=2\epsilon_0 c n_0 E_0^2$ and $\chi^{(3)}_{\mathrm{cw}}=4 n_0 \epsilon_0 c n_2 / 3$, this equation revrites: 
\begin{equation}
\langle\mathcal{P}^\mathrm{SPM}\rangle(t) = \frac{n_2 I }{n_0 c}\frac{3\omega_\mathrm{e}^2+5\omega^2}{\omega_\mathrm{e}^2-\omega^2} \frac{\mathrm{d}I}{\mathrm{d}t} 
\end{equation}

In typical conditions (e.g., at a wavelength of 800 nm), $\omega_\textrm{e} \gg \omega$, resulting in

\begin{equation}
\langle\mathcal{P}^\mathrm{SPM}\rangle(t) \approx \frac{3 n_2 I }{n_0 c} \frac{\mathrm{d}I}{\mathrm{d}t},
\label{P_moyen}
\end{equation}

On the leading edge of the pulse, ${\mathrm{d}I}/{\mathrm{d}t}>0$ so that $\langle\mathcal{P}^\mathrm{SPM}\rangle\ >0$: The field transfers energy to the medium and initiates the dipole oscillation, while on the trailing edge the dipoles return to rest and release their energy into the electromagnetic field. The net energy loss by the pulse is $\int_{-\infty}^{\infty}{\langle\mathcal{P}^\mathrm{SPM}\rangle\mathrm{d}t}=0$ since $I=0$ at both $t=\pm\infty$.
The energy storage in the propagation medium is therefore transient and results in a net energy transfer from the red-shifted leading edge of the pulse towards its blue-shifted trail. It therefore reconciles red- and blue-shifts with the simultaneous conservation of energy and of the photon number: A temporal slice of the pulse cannot be considered as an isolated system. Rather, the completion of the system requires either to consider the propagation medium together with the pulse.
For typical ultrashort laser filaments in air~\cite{KasparianScience,ChinHLLTABKKS05,BergeSNKW07,CouaironM07,KasparianW08} the relative energy transfer can reach \SI{1}{\%/cm}, enabling the strong reshaping occurring during their propagation.

An alternative way to understand the transient energy storage in the medium consists in grouping the dephasing $\phi^\mathrm{SPM}$ with the third-order susceptibility $\chi^{(3)}$ and the associated non-linear refractive index $n_2$, resulting in the effective values

\begin{eqnarray}
\chi^{\mathrm{(SPM)}}_\mathrm{eff} &=& \chi^{(3)} e^{i\phi^\mathrm{SPM}} \\
n_{2,\mathrm{eff}}^\mathrm{SPM} &=& n_2 e^{i\phi^\mathrm{SPM}}
\end{eqnarray}

Their imaginary components $\chi^{(3)} \sin{\phi^\mathrm{SPM}} \approx  \chi^{(3)} {\phi^\mathrm{SPM}}$ and $n_2 \sin{\phi^\mathrm{SPM}} \approx  n_2 {\phi^\mathrm{SPM}}$ are intrinsically associated with gain or loss.

\subsection{Link with self-steepening.}
The above-derived transient energy storage has to translate into a depletion of the pulse intensity in the front, and a growth in its trail. Indeed, let us consider the self-steepening term affecting the envelope~\cite{BergeSNKW07}:
\begin{equation}
\frac{\mathrm{d}\mathcal{E}}{\mathrm{d}z}=\frac{-n_2}{n_0 c}\frac{\mathrm{d}}{\mathrm{d}t}\left(|\mathcal{E}|^2\mathcal{E}\right)
\end{equation}
where $|\mathcal{E}|^2 = I$. The relative variation of $\mathcal{E}$ is therefore:
\begin{equation}
\frac{\mathrm{d}\mathcal{E}}{\mathcal{E}dz}=\frac{-n_2|\mathcal{E}|^2}{n_0 c}\left(\frac{\mathrm{d}|\mathcal{E}|^2}{|\mathcal{E}|^2\mathrm{d}t}+\frac{\mathrm{d}\mathcal{E}}{\mathcal{E}\mathrm{d}t}\right)
\end{equation}
which can convert into the relative variation of the intensity:
\begin{equation}
\frac{\mathrm{d}I}{2Idz}=\frac{-n_2 I}{n_0 c}\left(\frac{\mathrm{d}I}{I\mathrm{d}t}+\frac{\mathrm{d}I}{2I\mathrm{d}t}\right)
\end{equation}
The local intensity variations due to self-steepening therefore amount to
\begin{equation}
\frac{\mathrm{d}I}{\mathrm{d}z}=\frac{-3n_2 I}{n_0 c}\frac{\mathrm{d}I}{\mathrm{d}t}
\end{equation}
which identifies with the instantaneous power gained from the medium, $-\langle\mathcal{P}^\mathrm{SPM}\rangle$ (Eq.~(\ref{P_moyen})), providing an interpretation of the self-steepening in terms of the deformation of the envelope due to the conservation of the photon number density in spite of their energy drift due to the SPM-induced frequency change. This interpretation is fully compatible with the usual one in terms of the  stretching of the temporal pulse slices due to gradients in the group velocity. The latter focuses on the point of view of the wave deformation, while the former translates this deformation and the associated frequency shifts into photon energy and considers its implications for the simultaneous conservation of energy and the photon number.

\subsection{Time-dependent frequency shift in FWM.} 
We shall now highlight how the FWM formalism accounts for a time-dependent frequency shift in spite of the energy conservation.
We define the instantaneous spectrum at time $t_0$ as the Fourier transform of the pulse convolved by a temporal gate centered at $t_0$ and of width $\tau$ such that $ 1/\omega \ll \tau \ll T$, where $T$ is the pulse duration. To allow analytical derivations, we will consider a Gaussian gate in the following. However, other gate shapes may be considered without loss of generality.

Four-wave mixing is typically described by assuming an instantaneous response function of the third-order non-linear susceptibility (\Cref{eq:P3})
and focusing on the nonlinear polarization component in $EE^*E$, oscillating at $\omega$:
\begin{equation}
P^\mathrm{SPM}(t)= 4 \epsilon_0^2 c n_0 n_2 E_0(t)^3 \left(e^{-\mathrm{i}\omega t}+\mathrm{c.c.}\right)
\label{eqn:P_NL_chi}
\end{equation}
Under the paraxial approximation, this non-linear polarization contributes to the field evolution:
\begin{equation}
\left.2\mathrm{i} k_0 \frac{\mathrm{d} E(t)}{\mathrm{d} z}\right|_\mathrm{NL,\omega}=-\frac{\omega^2}{\epsilon_0 c^2} P^\mathrm{SPM}(t) \label{eqn:evo_parax}
\end{equation}
$k_0$ being the wavevector in vacuum. After plugging~(\ref{eqn:P_NL_chi}) into~(\ref{eqn:evo_parax}), the field evolves as
\begin{equation}
\frac{\mathrm{d} E(t)}{\mathrm{d} z}=\mathrm{i} C_\mathrm{NL} E(t)^3,
\end{equation}
where $C_\mathrm{NL}=2k_0 \epsilon_0 c n_0 n_2$. 
In order to evaluate the frequency shift of the instantaneous spectrum after a short propagation distance $\mathrm{d}z$, we develop the electrical field at the first order in $\mathrm{d}z$ as
\begin{equation}
E(t,\mathrm{d}z)=E(t,0)+\mathrm{i}C_\mathrm{NL}E(t,0)^3\mathrm{d}z + \mathcal{O}(\mathrm{d}z).
\label{eqn:propag_dz}
\end{equation}
Now let us locally Fourier-transform Eq.~(\ref{eqn:propag_dz}) in the vicinity of $t_0$ after Taylor-expanding the slowly-varying field envelope as $E_0(t)\approx E_0(t_0) +(t-t_0) E_0^\prime(t_0)$. Introducing $\Delta \omega = \bar{\omega} - \omega$, and neglecting the influence of dephasing on the field amplitude, we get the intensity spectrum at first order in $\mathrm{d}z$
\begin{equation}
I(\bar{\omega} \propto \left|\hat{E}(\bar{\omega}),t_0,\mathrm{d}z)\right|^2 \approx \mathrm{e}^{-\frac{\tau^2}{2}\Delta \omega^2} \tau^2 E_0^2(t_0) 
\times\left(\frac{1}{2}-3C_\mathrm{NL} E_0(t_0)E_0^\prime(t_0) \Delta \omega \mathrm{d}z\right),
\end{equation}
where all terms involving ${E_0^\prime}^2$ have been discarded based on the SVEA approximation. Deriving with respect to $\Delta \omega$ and Taylor-expanding again at order~$1$ in the vicinity of $\omega$ yields the spectral peak of the spectrum at time $t_0$:
\begin{eqnarray}
\Delta \omega^\ast (t_0) &=& -6 C_\mathrm{NL} E_0 E_0^\prime \mathrm{d}z
 \\
&=& - k_0 n_2 I_0'(t_0) \textrm{d}z \label{eqn:result_gen} 
\end{eqnarray}
where we have introduced the intensity~$I_0=2n_0 \epsilon_0 c E_0^2$.
Similarly, the mean frequency is, at first order in $\mathrm{d}z$:
\begin{eqnarray}
\langle \omega \rangle (t_0) \equiv \frac{\int \omega |\hat{E}|^2 \mathrm{d}\omega}{\int |\hat{E}|^2 \mathrm{d}\omega} 
&=&\omega-\frac{24 C_\mathrm{NL}E_0^3(t_0)E_0^\prime(t_0)\mathrm{d}z}{4E_0^2(t_0)+{E_0^\prime}^2\tau^2} \\
&\approx& \omega - 6 C_\mathrm{NL} E_0 E_0^\prime \mathrm{d}z \\
&=& \omega - k_0 n_2 I'(t_0) \textrm{d}z \label{Delta_omega}
\end{eqnarray}
where ${E_0^\prime}^2\tau^2 \ll E_0^2$ due to the SVEA.
Therefore, both the peak (Eq.~(\ref{eqn:result_gen})) and mean (Eq. (\ref{Delta_omega})) transient frequency offset calculated with the FWM formalism are equal to those predicted by SPM (Eq. (\ref{SPM})), provided the temporal intensity variation of the incident pulse is taken into account.
Note that the frequency offset does not depend on the width $\tau$ of the gate, which confirms that it is only a computation intermediate. 

\section{Conclusion}
As a conclusion, we reconciled the time-dependent frequency shift described by self-phase modulation, with the energy and photon number conservations implied by the four-wave mixing formalism. 
An energy transfer occurs from the red-shifted leading edge of the pulse to the blue-shifted trailing edge, mediated by a transient energy storage in the propagation medium. The latter can be evidenced by considering the dephasing between the driving pulse and the medium polarization, induced by the Kerr effect. This energy transfer is also the origin of self-steepening, which translates as a depletion (resp. replenishment) of the pulse envelope on its front (resp. trailing edge).
Furthermore, considering the time-dependent intensity in the FWM formalism reproduces this frequency shift, i.e., this transient energy storage in the medium.
Our results straightforwardly generalize to non-degenerate FWM and XPM, although the calculation is slightly more tedious. We also note that we did not consider group velocity dispersion (GVD). Taking it into account would affect FWM via the phase matching conditions, and equivalently SPM via the envelope deformation, without impact on our argument.

\textbf{Acknowledgements}. This work was supported by the Swiss NSF (contracts 200021\_155970 and 200020\_175697). We gratefully acknowledge fruitful discussions with Jean-Pierre Wolf, Nicolas Berti and Wahb Ettoumi.

\bibliographystyle{unsrt}

\begin{thebibliography}{10}

\bibitem{AlfanoS70b}
R.R. Alfano and S.L. Shapiro, Emission in the region 4000 to 7000 \AA ~via four-photon coupling in glass. Physical Review Letters. \textbf{24}, 584(1970)

\bibitem{AlfanoS70a}
R.R. Alfano and S.L. Shapiro, Observation of Self-Phase Modulation and Small-Scale Filaments in Crystals and Glasses. Physical Review Letters. \textbf{24}, 592 (1970)

\bibitem{Zheltikov2002}
A. M. Zheltikov. Ultrashort light pulses in hollow waveguides, Physics-Uspekhi \textbf{45}, 687--718 (2002)

\bibitem{DudleyRMP}
J.M. Dudley, G. Genty, S. Coen, Supercontinuum generation in photonic crystal fiber, Reviews of Modern Physics \textbf{78}, 1135 (2006)

\bibitem{KnightS07}
J. C. Knight, D. V. Skryabin, Nonlinear waveguide optics and photonic crystal fibers, Optics Express \textbf{15}, 15365 (2007)

\bibitem{Alfano06pXX}
R. R. Alfano, The Supercontinuum laser source, Fundamentals with updated references, Second Edition, Springer, 2006, pp. 48, 127, 132, 135, 524

\bibitem{BoydXX}
R. W. Boyd, "Non-linear Optics", Academic Press, New York, 2001, page 3 

\bibitem{He_Liu}
Guang S. He, Song H. Liu, Physics of nonlinear optics, World Scientific, Singapore, 1999, p. 81

\bibitem{LoudonXX}
Rodney Loudon, Quantum theory of light, 3rd edition, Oxford Science Publications, Oxford, UK, (2000) p. 411

\bibitem{Agrawal}
Govind P. Agrawal Nonlinear Fiber Optics, Academic Press, 2001, p. 101

\bibitem{Shen}
Y. R. Shen, "The Principles of Nonlinear Optics", Wiley, 2003, p. 328

\bibitem{StricklandM85}
D. Strickland, G. Mourou
\newblock "Compression of amplified chirped optical pulses"
\newblock{Opt. Commun.} \textbf{56}, 219-221 (1985)


\bibitem{EttoumiPKW10}
W. Ettoumi, Y. Petit, J. Kasparian, J.-P. Wolf, 
Generalized Miller formul\ae, 
Opt. Express \textbf{18}, 6613-6620 (2010)


\bibitem{ZhangLW2008}
J. Zhang, Z. H. Lu,  L. J. Wang, 
Precision refractive index measurements of air, N$_2$, O$_2$, Ar, and CO$_2$ with a
frequency comb, 
Appl. Opt. \textbf{47}, 3143 (2008)

\bibitem{KasparianW08}
J.~Kasparian and J.-P. Wolf.
\newblock Physics and applications of atmospheric non-linear optics and filamentation.
\newblock {Opt. Express} \textbf{16}, 466-493, (2008).

\bibitem{BergeSNKW07}
L.~Berg\'e, S.~Skupin, R.~Nuter, J.~Kasparian, and J.-P. Wolf.
\newblock Ultrashort filaments of light in weakly ionized, optically transparent media.
\newblock {Rep. Prog. Phys.} \textbf{70}, 1633-1713, (2007).

\bibitem{CouaironM07}
A.~Couairon and A.~Mysyrowicz.
\newblock Femtosecond filamentation in transparent media.
\newblock {Phys. Rep.} \textbf{441}, 47-189, (2007).

\bibitem{ChinHLLTABKKS05}
S. L. Chin, S. A. Hosseini, W. Liu, Q. Luo, F. Theberge, N. Akozbek, A. Becker, V. P. Kandidov, O. G. Kosareva, and H. Schroeder.
\newblock The propagation of powerful femtosecond laser pulses in optical media: physics, applications, and new challenges.
\newblock {Can. J. Phys.} \textbf{83}, 863-905 (2005).

\bibitem{KasparianScience}
J. Kasparian, M. Rodriguez, G. M\'ejean, J. Yu, E. Salmon, H. Wille, R. Bourayou, S. Frey, Y.-B. Andr\'e, A.~Mysyrowicz, R. Sauerbrey, J.-P. Wolf, L. W\"oste.
\newblock White-Light Filaments for Atmospheric Analysis.
\newblock {Science} \textbf{301}, 61-64 (2003)


\end{thebibliography}

\end{document}